\documentclass{PoS}

\usepackage{amsmath}
\usepackage{braket}
\usepackage{cite}

\title{Pad\'e approach to top-quark mass effects in gluon fusion amplitudes}

\ShortTitle{Pad\'e approach to top-quark mass effects}

\author{Joshua Davies,$^{a}$ Ramona Gr\"ober,$^{b}$ Andreas Maier,$^{c}$ \speaker{Thomas Rauh}$^{d}$
and Matthias Steinhauser$^{a}$\\
\llap{$^a$} Institut f \"ur Theoretische Teilchenphysik, \\
   Karlsruhe Institute of Technology (KIT),\\
   Wolfgang-Gaede Stra{\ss}e 1, 76128 Karlsruhe, Germany\\
\llap{$^b$} Dipartimento di Fisica e Astronomia ``G. Galilei'', Universit\`a di Padova, and \\
   Istituto Nazionale di Fisica Nucleare, Sezione di Padova,\\
   I-35131 Padova, Italy\\
\llap{$^c$} Deutsches Elektronen-Synchrotron, DESY, \\
   Platanenallee 6, D-15738 Zeuthen, Germany\\
\llap{$^d$} Albert Einstein Center for Fundamental Physics,\\
   Institute for Theoretical Physics, University of Bern,\\
   Sidlerstrasse 5, CH-3012 Bern, Switzerland\\
E-mail: \email{ joshua.davies@kit.edu}, \email{ramona.groeber@pd.infn.it}, \email{andreas.martin.maier@desy.de}, 
\email{rauh@itp.unibe.ch}, \email{matthias.steinhauser@kit.edu}}

\abstract{Gluon fusion processes like single and double Higgs production exhibit 
slow convergence and pose severe computational challenges. We show how the 
top-quark mass dependence of the virtual amplitudes can be reconstructed with a 
conformal mapping and Pad\'e approximants based on the expansion in the inverse 
top-quark mass and the non-analytic terms in the expansion around the top threshold. 
The method is then applied at two- and three-loop order.}

\FullConference{14th International Symposium on Radiative Corrections (RADCOR2019)\\ 
9-13 September 2019\\
		Palais des Papes, Avignon, France}

\dedicated{TTP19-044\\ P3H-19-049\\ DESY 19-223\\ SAGEX-19-33}

\begin{document}


\section{Introduction\label{sec:intro}}

\begin{figure}[b]
\begin{center}
\begin{tabular}{ccc}
\includegraphics[height=1.8cm]{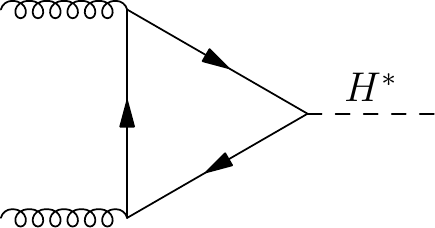}&
\includegraphics[height=1.8cm]{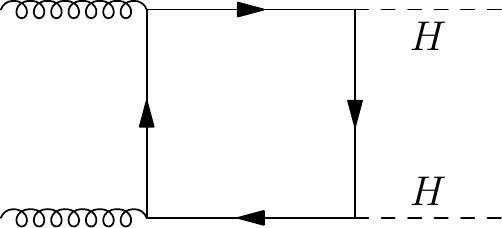}&
\includegraphics[height=1.8cm]{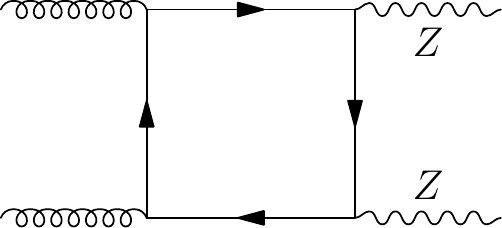}\\[0.5cm]
\includegraphics[height=1.8cm]{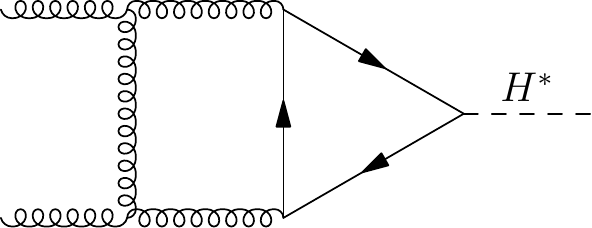}&
\includegraphics[height=1.8cm]{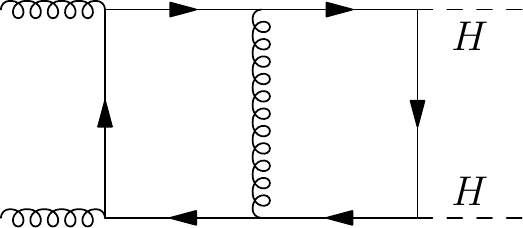}&
\includegraphics[height=1.8cm]{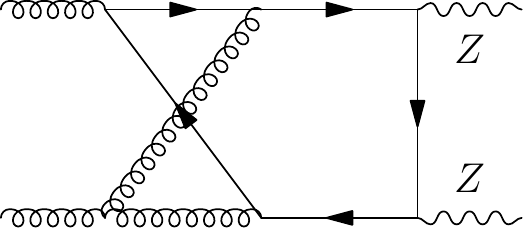}
\end{tabular}
\end{center}
\caption{Sample diagrams for gluon fusion processes at LO (top) and NLO (bottom). \label{fig:diagrams}}
\end{figure}

Following the advent of precision Higgs physics, top-quark mass effects are currently a 
very active topic, which was also heavily featured at this conference~\cite{ProcRadcor}. 
We show sample diagrams for some of the relevant processes in Figure~\ref{fig:diagrams}, 
where the fermion lines correspond to the top quark. 
These gluon-fusion processes have in common that they are loop-induced and that their 
perturbative expansions only converge slowly, implying that higher-order corrections 
are both important and very challenging to compute. Traditionally, the infinite top-quark 
mass limit has been used beyond leading order to considerably simplify calculations. 
In many cases, higher powers in the large top-quark mass expansion (LME) were computed to 
refine predictions and obtain rough estimates for the uncertainties related to the 
approximation. While the expansion yields reliable results for single on-shell Higgs production, 
it breaks down in kinematic regimes of $2\to2$ processes where the internal top quarks 
can be on their mass shell, e.g. at large invariant masses or transverse momenta. 

In recent years numerical NLO results have become available \cite{Borowka:2016ehy,Borowka:2016ypz,Jones:2018hbb,Baglio:2018lrj,Maltoni:2018zvp,Chen:2019fla}, 
but these calculations require a high degree of optimization and still pose serious challenges. 
Therefore, there has also been considerable recent interest in expansion methods which are viable 
in an extended or complementary part of the phase space \cite{Bonciani:2018omm,Davies:2018ood,Davies:2018qvx,Xu:2018eos}. 
Here, we report on a different approach \cite{Grober:2017uho} based on the reconstruction 
of the amplitude from the LME and the expansion around the top-quark threshold $\hat{s}=4m_t^2$ 
with a conformal mapping and Pad\'e approximants~\cite{Baikov:1995ui,Broadhurst:1993mw,Fleischer:1994ef}. 
This method has the advantage that it is applicable in the entire physical phase space and 
that calculations are manageable at three-loop order where the other expansion methods are 
not feasible with the current computational technology. We first review our approach for 
the simpler case of single-Higgs production \cite{Davies:2019nhm} and then discuss 
double-Higgs production \cite{Grober:2017uho} and the Higgs-interference contribution 
to double-$Z$ production \cite{Grober:2019kuf}.


\section{Single-Higgs production\label{sec:ggH}}

\begin{figure}[t]
\begin{center}
 $\vcenter{\hbox{\includegraphics[height=5cm]{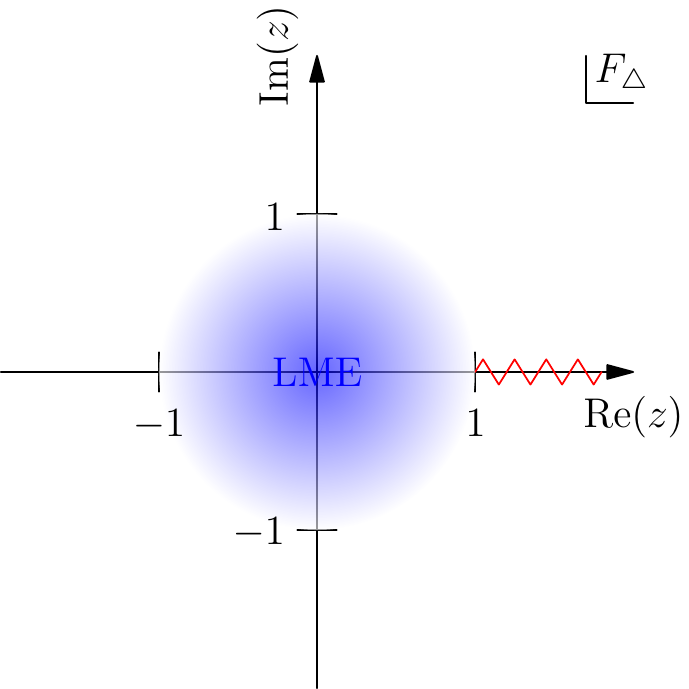}}}$
 \hspace{0.5cm}{\Large $\xrightarrow{z \, = \, \frac{4\omega}{(1+\omega)^2}}$}\hspace{0.5cm}
 $\vcenter{\hbox{\includegraphics[height=5cm]{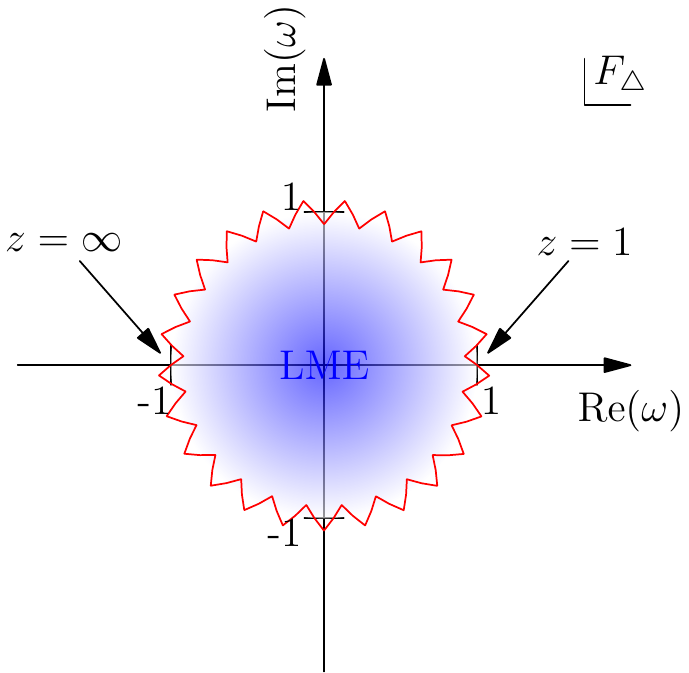}}}$
\end{center}
\caption{Analytic structure of the triangle form factor in the complex $z$ and $\omega$ planes. 
The red zigzag curve denotes the $t\bar{t}X$ branch cut and the blue shaded region illustrates 
the convergence properties of the LME. \label{fig:z_w_plane}}
\end{figure}

The amplitude for the production of an off-shell Higgs boson in gluon fusion 
\begin{equation}
 \mathcal{A}^{\mu\nu}_{ab}(g(p_1,\mu,a),g(p_2,\nu,b)\to H^{(*)}(p_H)) = \delta_{ab}T_F\frac{y_t\hat{s}}{\sqrt{2}m_t}\frac{\alpha_s}{4\pi}\left(g^{\mu\nu}-\frac{p_1^\nu p_2^\mu}{p_1\cdot p_2}\right)F_\triangle(z)
 \label{eq:AggH_def}
\end{equation}
only depends on a single form factor $F_\Delta(z)$ that is a function of the dimensionless ratio 
$z=\hat{s}/(4m_t^2)$, where $\hat{s} = (p_1+p_2)^2$ is the partonic center-of-mass energy. 
We apply the IR subtractions determined in~\cite{Catani:1998bh} to define finite form factors 
$F_\Delta^{(1),\text{fin}}$ and $F_\Delta^{(2),\text{fin}}$ at two and three-loop order, respectively. 
Causality implies that the form factor is an analytic function in the entire complex $z$-plane 
apart from a branch cut starting at the top-quark threshold $z=1$ as illustrated in 
Figure~\ref{fig:z_w_plane}, which limits the radius of convergence of the LME to unity. Beyond 
leading order there is also a branch cut from massless intermediate states (e.g. from cutting 
the gluon loop in the bottom-left diagram in Figure~\ref{fig:diagrams}) starting at $z=0$ which 
we will discuss below. 
We now apply the conformal transformation 
\begin{equation}
 z = \frac{4\omega}{(1+\omega)^2}\,,
 \label{eq:conf_trafo}
\end{equation}
which maps the complex $z$-plane onto the unit disc $|\omega|\leq1$ and the branch cut on the 
perimeter as shown in Figure~\ref{fig:z_w_plane}. This implies that the form factor can be 
approximated with a power series in $\omega(z)$ by determining the coefficients from the condition 
that the expansion of the ansatz in $z$ must reproduce the LME. A more general approach is to 
use Pad\'e approximants 
\begin{equation}
  [n/m](\omega) = \frac{\sum\limits_{i=0}^n a_i \omega^i}{1 + \sum\limits_{j=1}^m b_j \omega^j}\,,
  \label{eq:Pade_ansatz}
\end{equation}
which perform best when the polynomial degrees of the numerator and denominator are similar $n\approx m$. 
While it is possible to construct Pad\'e approximants with only the LME as an input to determine 
the coefficients $a_i$ and $b_j$, the quality of the results deteriorates with increasing loop order. 
This can be understood from the expansion of the form factor around the top threshold 
\begin{equation}
 F_\Delta^{(N),\text{fin}} \,\,\mathop{=}\limits^{z\to1}\,\, \sum\limits_{n=0}^{N} \left[
    \sum\limits_{i=0}^\infty A_{in} \bar{z}^i + 
    \sum\limits_{i=3-N}^\infty \, \sum\limits_{j=i\,\text{mod}\,2}^N  B_{ijn} \bar{z}^{\,i/2} \ln^j(\bar{z})
    \right]\ln^n(-4z+i0)\,,
 \label{eq:THR}
\end{equation}
where $\bar{z}=1-z$ and the $A_{in}$ and $B_{ijn}$ are coefficients. Again, let us first ignore 
massless cuts and focus on the contribution with $n=0$. In addition to a power series in $\bar{z}$ 
there are non-analytic contributions involving square roots and logarithms of $\bar{z}$. The 
leading power at which these terms appear depends on the loop order due to the $\alpha_s/v\sim\alpha_s/\sqrt{\bar{z}}$ 
scaling of Coulomb singularities near the threshold. While the roots are tamed by the conformal 
mapping $\sqrt{\bar{z}} = (1-\omega(z))/(1+\omega(z))$, which is analytic near $\omega=1$, the logarithms 
$\ln(\bar{z}) = 2[\ln(1-\omega(z)) - \ln(1+\omega(z))]$ constitute non-analytic terms near $\omega=1$ which 
cannot be reproduced by the Pad\'e ansatz \eqref{eq:Pade_ansatz}. The appearance of higher powers of 
these logarithms with increasing loop orders thus explains the worse behaviour of Pad\'e approximants 
constructed only from the LME. 

\begin{figure}[t]
\begin{center}
 \includegraphics[width=0.49\textwidth]{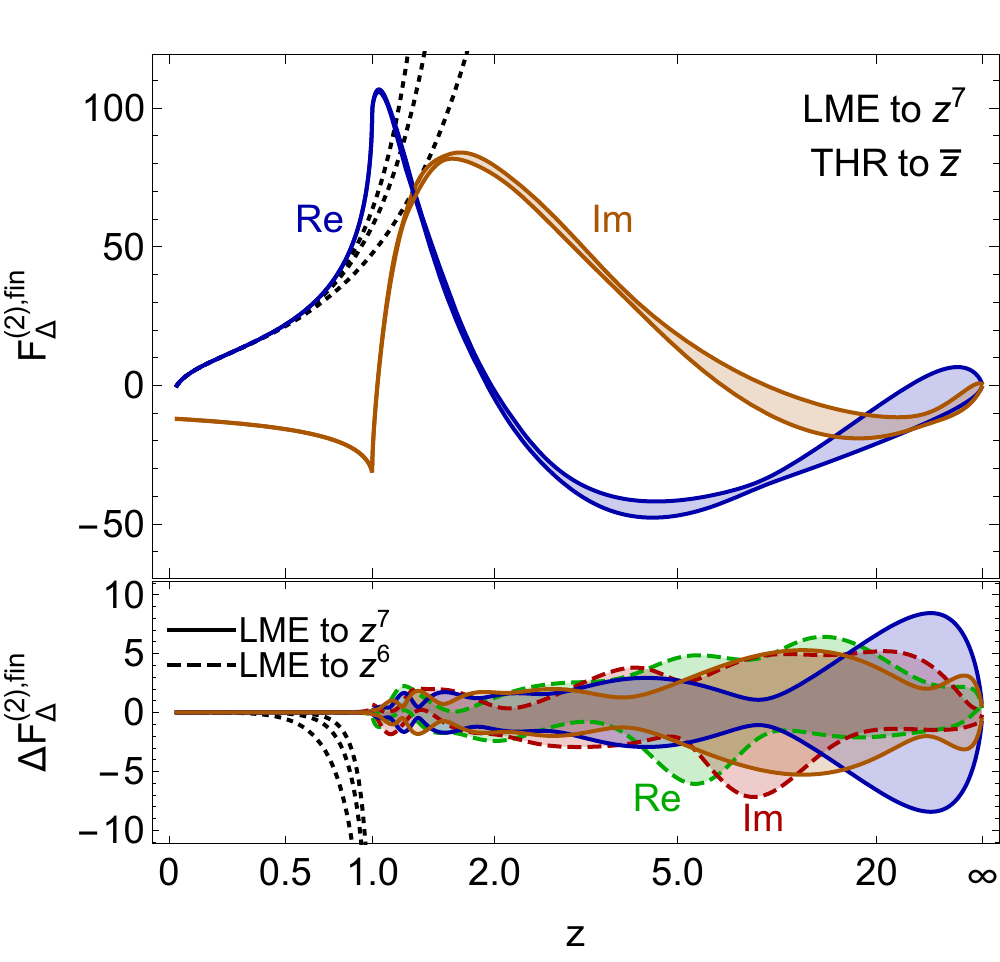} \hfill
 \includegraphics[width=0.5\textwidth]{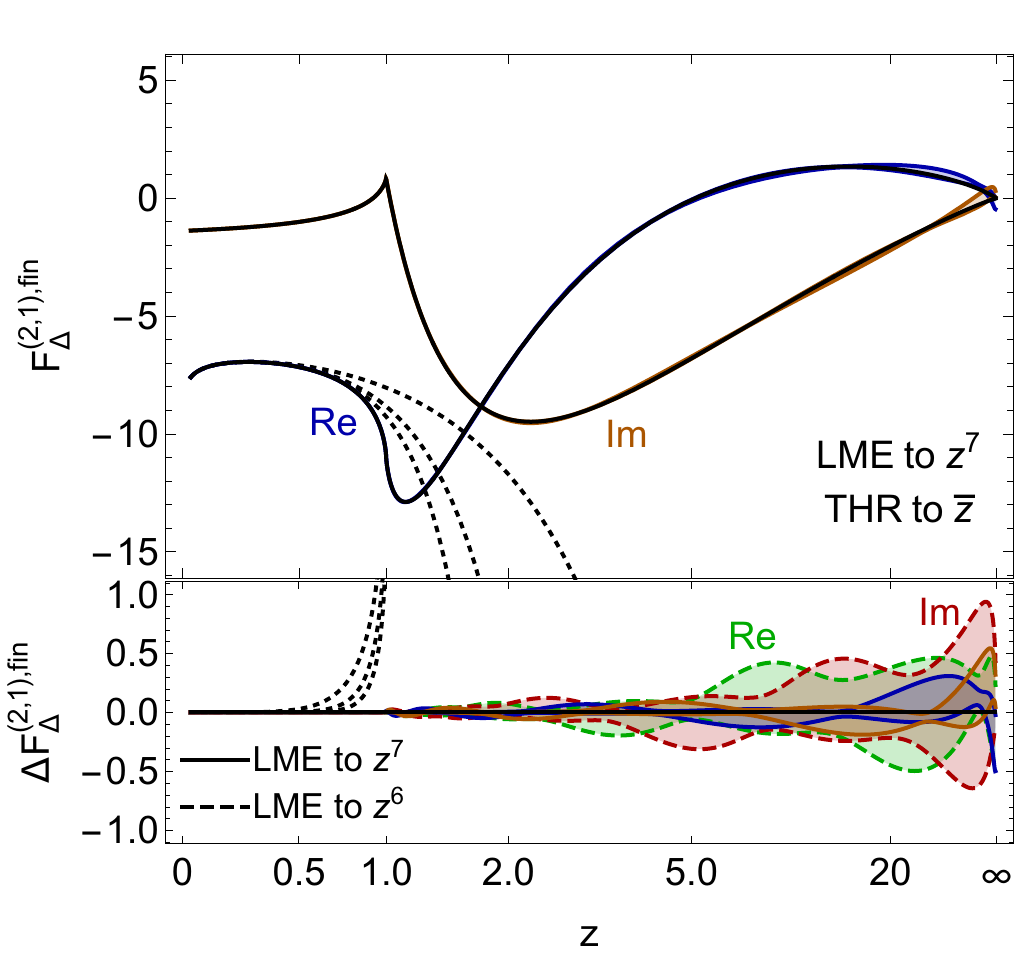}
\end{center}
\caption{The left plot shows our reconstructed results for the three-loop form factor $F_\Delta^{(2),\text{fin}}$ and 
the right plot compares our approximation of the fermionic contribution $F_\Delta^{(2,1),\text{fin}}$ to the analytic 
calculation (black) from~\cite{Harlander:2019ioe}. The blue and yellow regions correspond to our uncertainty 
estimate for the real and imaginary part of the form factor. In the lower panel we normalize the result to the 
central value (left) or the exact result (right) and furthermore show the approximation from~\cite{Davies:2019nhm} 
where one fewer LME coefficient was known in green and red. \label{fig:Ftri3}}
\end{figure}

To improve the reconstruction of the form factor we computed the non-analytic part of the threshold 
expansion \eqref{eq:THR}. For example, for $N=1$ (two loops) we compute $B_{ij0}$ with $j=1$ for 
$i=2,4,6$ and $j=0,1$ for $i=3,5$.\footnote{At two-loop order the terms proportional to $\ln(-4z+i0)$ are 
removed by the IR subtraction.} This is significantly simpler than the calculation of the analytic terms 
and sufficient to obtain a good reconstruction of the form factor at least up to three-loop order. 
The calculation is described in \cite{Grober:2017uho} and utilizes the non-relativistic effective 
field theory methods developed for the computation of the $e^+e^-\to t\bar{t}X$ cross section near the 
top-pair production threshold~\cite{Beneke:2015kwa,Beneke:2013kia,Beneke:2015lwa,Beneke:2016kkb,Beneke:2017rdn}. 
Our reconstruction takes the form 
\begin{equation}
 F_\triangle^{(2),\text{fin}}(z) \simeq  \frac{[n/m]\left(\omega(z)\right)}{1 + a_{R,0}z} + \frac{[k/l]\left(\omega(z)\right)}{1 + a_{R,1}z}\,\ln(-4z+i0) + s(z)\,,
 \label{eq:Ftri_Pade}
\end{equation}
where the known logarithmic terms have been absorbed into the subtraction function $s(z)$. The remainder 
is then free of threshold logarithms $\ln(\bar{z})$ up to the computed order in the threshold expansion 
and better suited for Pad\'e approximants. Here, we separately reconstruct the coefficient of the logarithm 
$\ln(-4z)$ stemming from massless cuts. We use a rescaling with $1 + a_{R,i}\,z$ to enforce the scaling 
$F_\Delta\to0$ as $z\to\infty$ which is a consequence of unitarity. Furthermore, randomly choosing the 
parameters $a_{R,i}$ in the range $[0.1,10]$ allows us to construct multiple Pad\'e approximants and thus 
estimate the uncertainty of the results. In \cite{Davies:2019nhm} we presented our approximants for the 
three-loop form factor including input from the LME up to the order $z^6$ and the threshold expansion 
up to the order $\bar{z}$. 

After the appearance of our work the $z^7$ coefficient in the LME has been computed in \cite{Davies:2019djw} 
and the light-fermion contributions have been calculated analytically in \cite{Harlander:2019ioe}. 
In Figure~\ref{fig:Ftri3} we update our results from \cite{Davies:2019nhm} with Pad\'e approximants constructed 
using the additional LME coefficient and compare the fermionic contribution to the exact calculation. 
Sample Pad\'e approximants which include the new input are available from \cite{KIT}. 
We observe that the uncertainties are tiny below the top threshold and then moderately increase for 
larger values of $z$ while remaining sufficiently small for phenomenological applications throughout the 
entire range. The lower panel shows normalized results with and without the $z^7$ coefficient and we find 
good consistency between the two approximations. Our result for the light-fermion contribution 
shows very good agreement with the analytic calculation~\cite{Harlander:2019ioe} throughout the entire 
range and the uncertainty is reduced by the addition of the new LME coefficient. This demonstrates that the 
Pad\'e approximants can further be improved by including additional information from the kinematic expansions.


\section{\boldmath Pad\'e method for $gg\to AB$ processes\label{sec:2to2}}

\begin{figure}[t]
\begin{center}
 \includegraphics[width=0.9\textwidth]{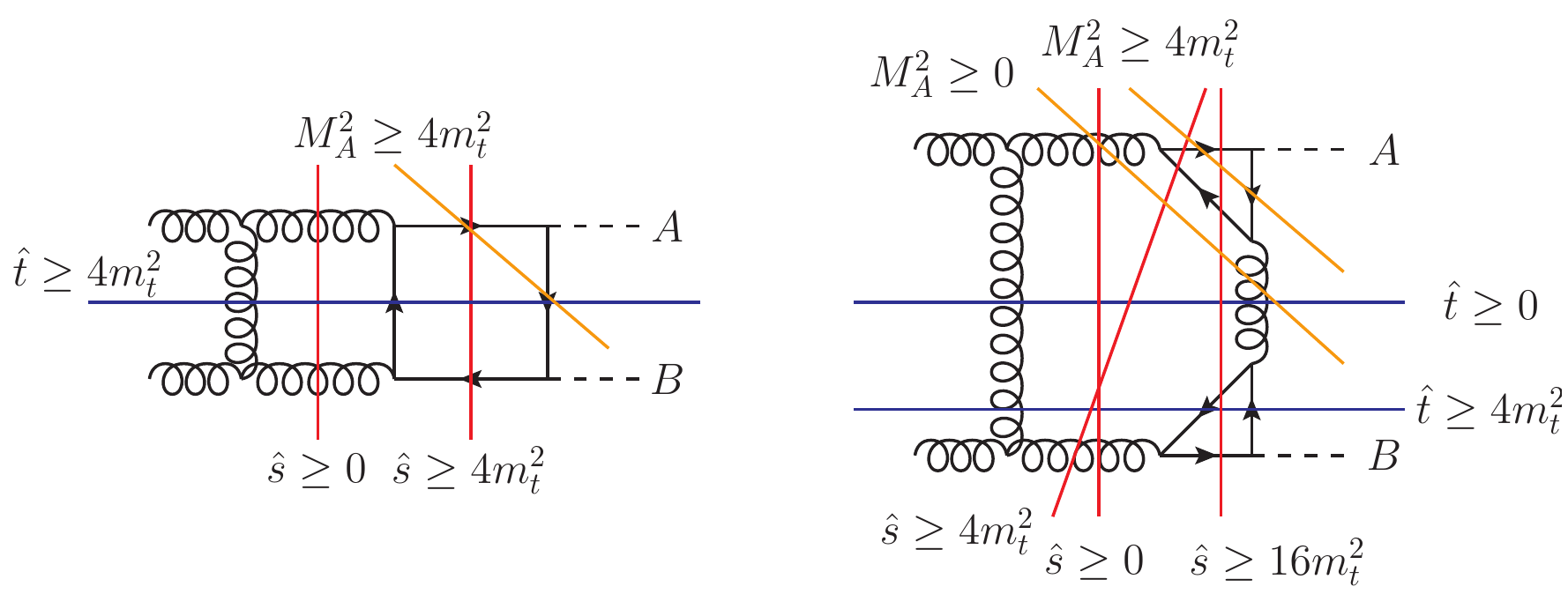}
\end{center}
\caption{Examples for on-shell cuts in two diagrams contributing to the $gg\to AB$ process.\label{fig:ggAB_cuts}}
\end{figure}

The method we described in Section~\ref{sec:ggH} can also be applied to the kinematically more 
complicated amplitudes of $gg\to AB$ scattering processes. First we decompose the amplitude into 
a number of form factors which will then be reconstructed. These form factors depend on several 
kinematic quantities: the invariant masses $M_A$, $M_B$ of the final state particles, the Mandelstam 
variables $\hat{s},\hat{t},\hat{u}$ which are subject to the condition 
\begin{equation}
 \hat{s}+\hat{t}+\hat{u} = M_A^2 + M_B^2\,,
 \label{eq:Mandelstam_condition}
\end{equation}
and the mass $m_t$ of the internal top quarks. This implies that the analytic structure of the form 
factors is much more complicated than for $gg\to H^*$. We show several on-shell cuts in sample diagrams 
in Figure~\ref{fig:ggAB_cuts}. We divide the form factors into coefficients of logarithms $\ln(-\hat{s}/m_t^2+i0)$,
$\ln(-M_A^2/m_t^2+i0)$, etc. from the massless cuts $\hat{s}\geq0$, $M_A^2\geq0$, etc. as done for the $\hat{s}$-channel 
cut in \eqref{eq:Ftri_Pade}. These coefficients are then reconstructed separately and we proceed as follows. 

First, we specify a phase space point by fixing the kinematic variables $\hat{s},\hat{t},\hat{u},M_A^2,M_B^2$ in the physical 
region. The variable $z$ then encodes the top-quark mass dependence of the amplitude at the given phase space 
point. The analytic structure is similar to that displayed on the left hand side in Figure~\ref{fig:z_w_plane}, 
but there are also branch cuts at $z\leq \hat{s}/\hat{t},\hat{s}/\hat{u}$, corresponding to negative values of $m_t^2$, from the $t$ 
and $\hat{u}$ channel. These values of $z$ map onto real negative values of $\omega$ inside the unit circle under 
the conformal transformation \eqref{eq:conf_trafo}. While the branch cuts cannot be reproduced by the Pad\'e 
ansatz, the negative values of $m_t^2$ are obviously unphysical and we observe that the leading-order form 
factors for $gg\to HH$ and the Higgs-interference contribution to $gg\to ZZ$ can be reproduced almost 
perfectly in spite of them \cite{Grober:2017uho,Grober:2019kuf}. A similar situation has been discussed in 
\cite{Masjuan:2009wy} where it was shown that the poles in the Pad\'e ansatz \eqref{eq:Pade_ansatz} accumulate 
in the region of the branch cut and thus approximate the cut. Since the Pad\'e approximants do not need to be 
evaluated in the vicinity of the $\hat{t}$ and $\hat{u}$ channel thresholds we conclude that the corresponding cuts do 
not impede the reconstruction of the amplitude with our method. The remaining branch cuts at $\hat{s}\geq16m_t^2$, 
$M_A^2\geq4m_t^2$ etc. lie on top of the $t\bar{t}X$ cut with threshold that are significantly larger than $z=1$ 
and do not pose any problems. 

\begin{figure}[t]
\begin{center}
 \mbox{\hspace{-0.5cm}\includegraphics[width=0.56\textwidth]{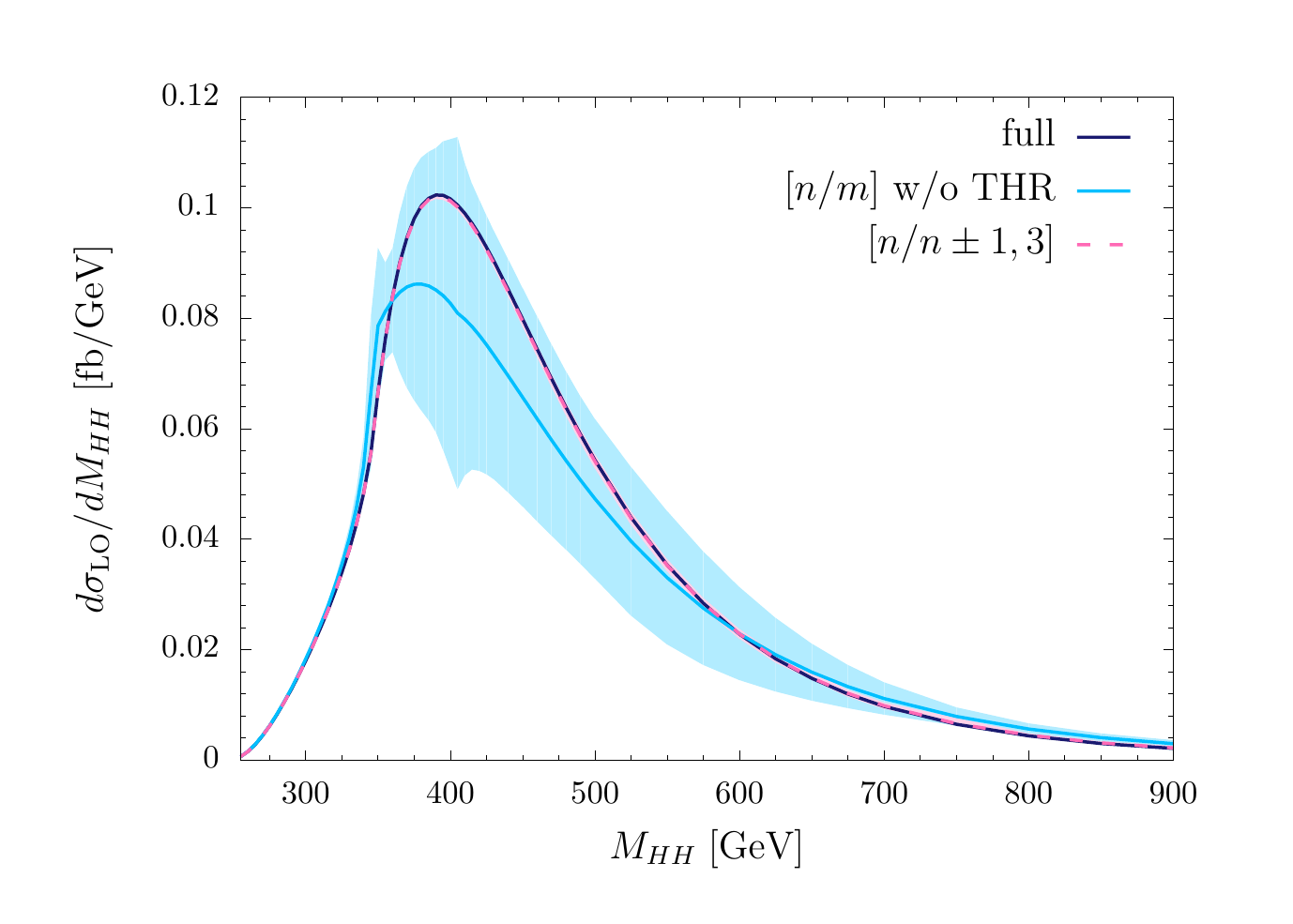}
 \hspace{-1cm}\includegraphics[width=0.56\textwidth]{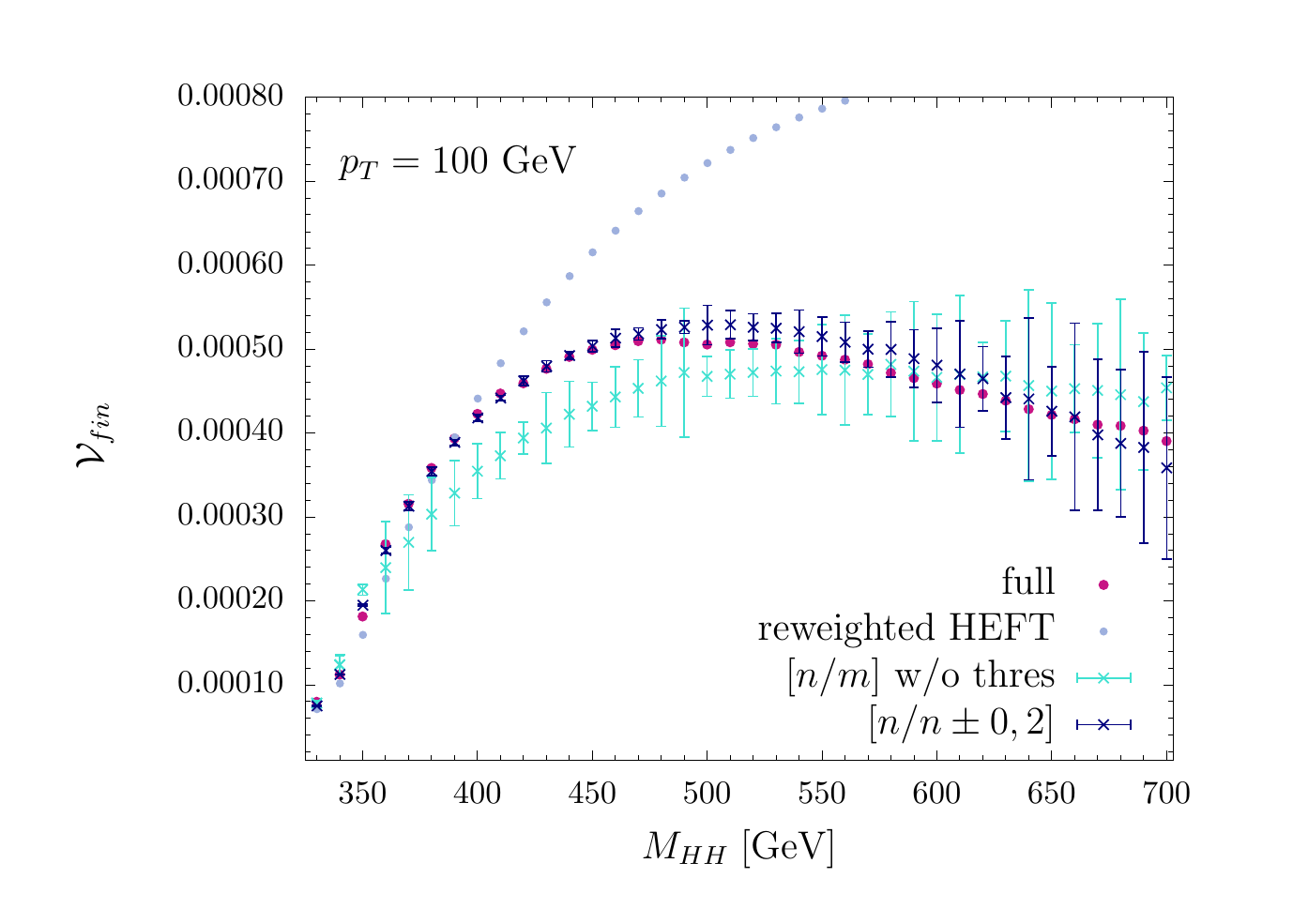}\hspace{-0.8cm}}
\end{center}
\caption{The left plot shows the leading-order invariant mass distribution of Higgs-pair production. 
The dark blue line is the exact result and the pink-dashed line corresponds to our full reconstruction 
where the uncertainty is too small to be seen. The light-blue line shows the reconstruction without 
input from the threshold expansion and the light-blue region corresponds to the uncertainty estimate. 
On the right we compare the IR-subtracted squared matrix element $\mathcal{V}_\text{fin}$ with the 
numerical calculation from \cite{Borowka:2016ehy,Borowka:2016ypz} (pink points). Our results with 
and without input from the top threshold are given in dark blue and light blue, respectively.\label{fig:ggHH}}
\end{figure}

Our approximate result follows by constructing Pad\'e approximants in the variable $\omega(z)$ and evaluating 
them for the physical value of $m_t$. We use random values of the rescaling parameters $a_{R,i}$ defined as 
in \eqref{eq:Ftri_Pade} to obtain a total of 100 $[n/m]$ approximants where we apply the condition $|n-m|\leq3$ 
since Pad\'e approximants with a similar polynomial degree in the numerator and denominator generally perform best. 
Furthermore, we exclude approximants with poles close to real positive values of $z$ in the complex plane. 
The standard deviation of the 100 values is used as an estimate of the uncertainty. These steps need to be 
performed for every phase space point which typically requires a few CPU seconds.

\subsection{Double-Higgs production}

Double-Higgs production is the best studied case of top-quark mass effects in gluon-fusion 
processes with two-particle final states where the first numerical two-loop calculations 
\cite{Borowka:2016ehy,Borowka:2016ypz} appeared. This allows a detailed comparison of our 
approach which is shown in Figure~\ref{fig:ggHH}. We constructed Pad\'e approximants from 
the LME up to the order $z^4$ \cite{Grigo:2015dia,Degrassi:2016vss} and the threshold 
expansion up to the order $\bar{z}^{\,5/2}$ (LO) and $\bar{z}^{\,2}$ (NLO) \cite{Grober:2017uho}. 
This provides an almost perfect reconstruction at LO with tiny uncertainties. The comparison 
with the approximation without input from the threshold expansion demonstrates the importance 
of that additional  information for the method. At NLO we find that the reconstructed result 
matches the numerical calculation very well, albeit with bigger errors that increase towards 
larger invariant masses. Larger uncertainties are expected because most of the complications 
discussed above, i.e. massless cuts and threshold logarithms, first appear at two-loop order. 
In particular we find that the coefficient of the logarithm $\ln(-\hat{s}/m_t^2+i0)$ from the massless 
$\hat{s}$-channel cut has a larger relative uncertainty than the non-logarithmic term. Still, the 
uncertainties are small enough for phenomenological applications and can be further reduced 
with more input from the LME and the threshold expansion.

\subsection{Higgs-interference contribution to $Z$-pair production}

\begin{figure}[t]
\begin{center}
 2\,Re\,$\left(\quad\vcenter{\hbox{\includegraphics[height=2cm]{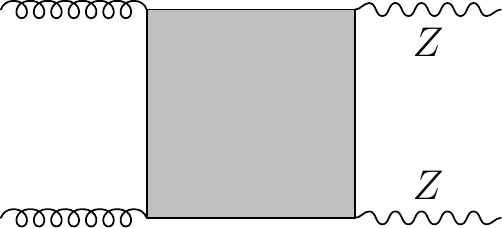}}}\right.$
 \hspace{0.5cm}{\Large $\times$}\hspace{0.5cm}
 $\left.\vcenter{\hbox{\includegraphics[height=2cm]{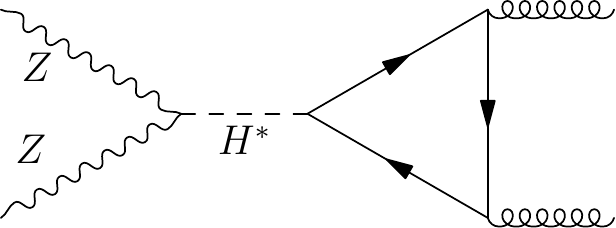}}}\quad\right)$
\end{center}
\caption{Off-shell Higgs interference contribution to $Z$-pair production in gluon fusion. \label{fig:ggZZinterference}}
\end{figure}

\begin{figure}[t]
\begin{center}
\includegraphics[width=0.49 \textwidth]{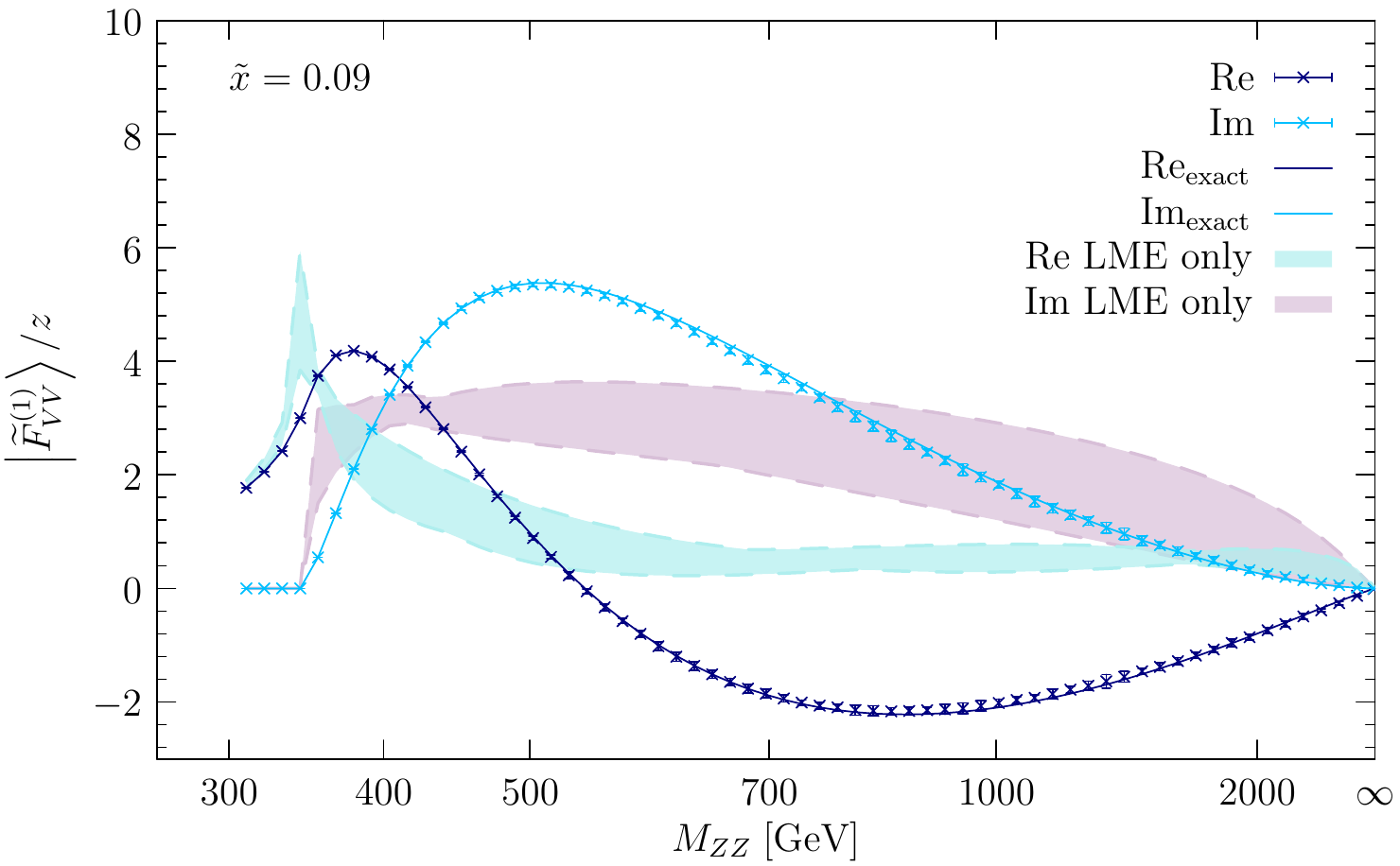}\hfill
\includegraphics[width=0.49 \textwidth]{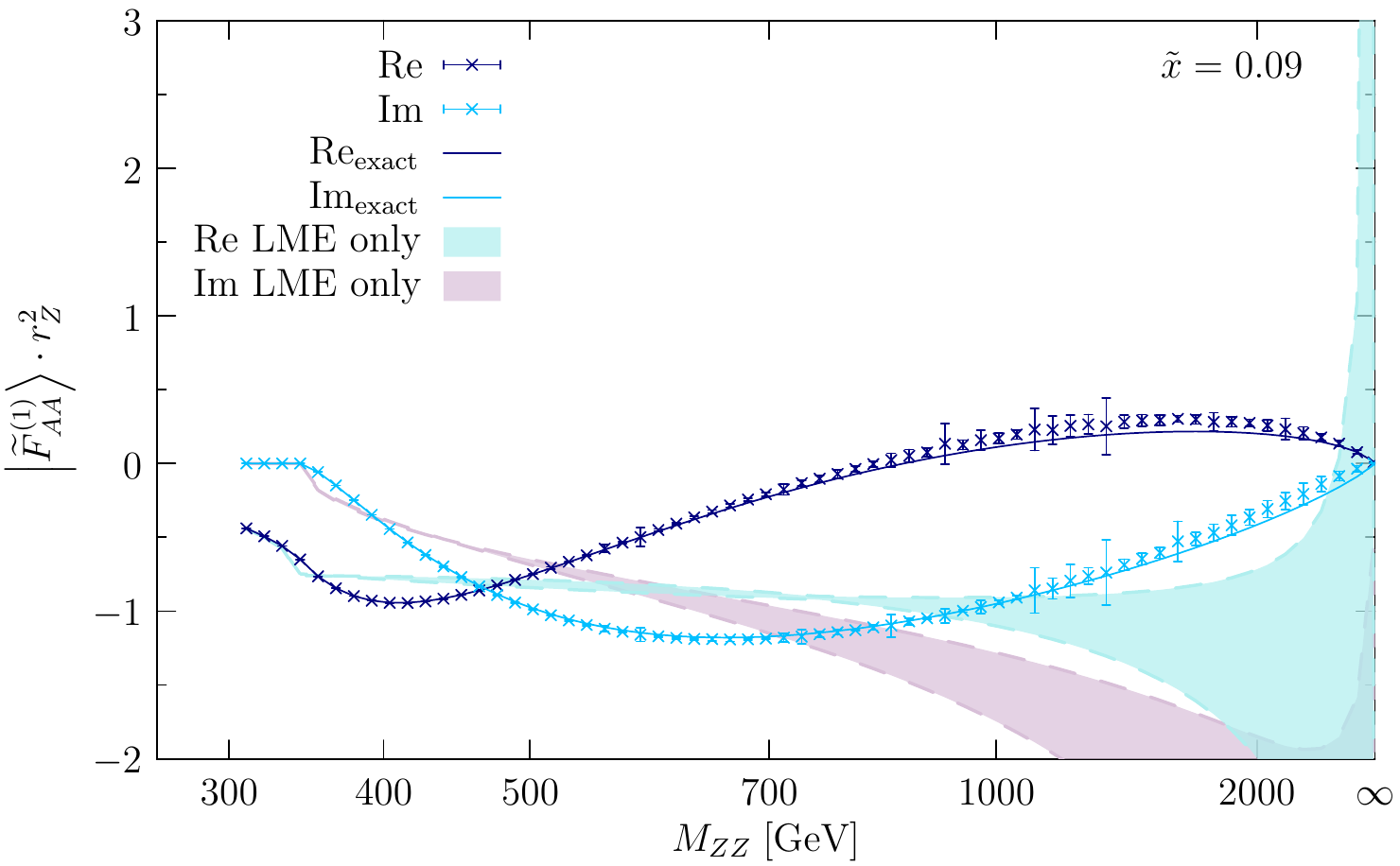}\\
\includegraphics[width=0.49 \textwidth]{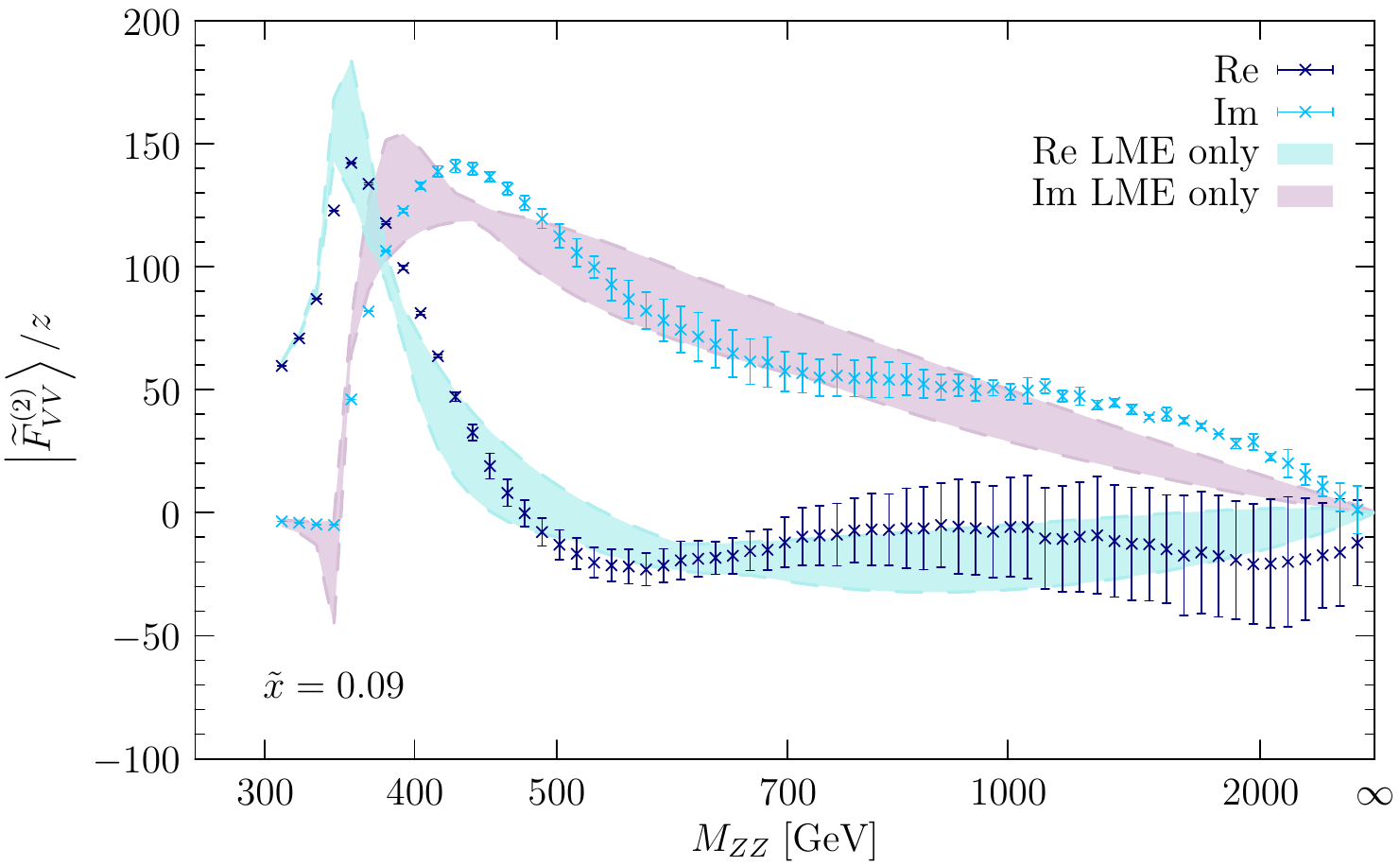}\hfill
\includegraphics[width=0.49 \textwidth]{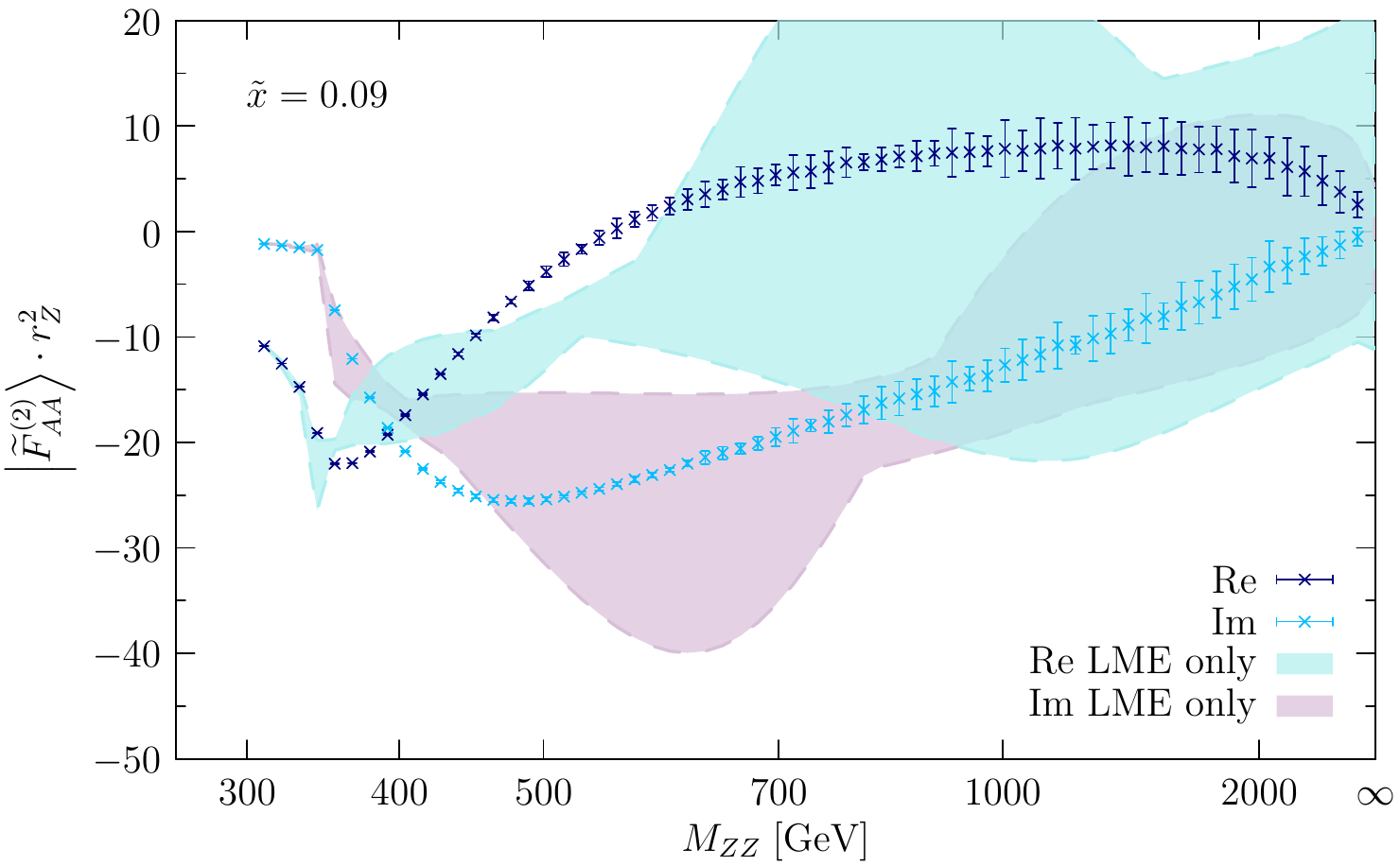}
\end{center}
\caption{The LO (top) and NLO (bottom) form factors $\Ket{\widetilde{F}_{VV}^{(i)}}$ (left) 
and $\Ket{\widetilde{F}_{AA}^{(i)}}$ are shown as a function of the invariant mass of the 
$Z$-boson pair at a fixed value of $\tilde{x} = (p_T^2+m_Z^2)/M_{ZZ}^2 = 0.09$. The real and 
imaginary parts of the Pad\'e approximants are shown in dark blue and light blue and the solid 
lines correspond to the exact result at LO. The shaded regions show an approximation constructed 
from the LME only, following the approach of \cite{Campbell:2016ivq}. \label{fig:ggZZ}}
\end{figure}

Measuring the process $gg\to H \to ZZ\,(\to4l)$ on and off the Higgs-boson peak provides an 
indirect way to measure the Higgs-boson width $\Gamma_H\sim4\,$MeV~\cite{Kauer:2012hd,Caola:2013yja,Campbell:2013una}, 
which is not directly accessible at the LHC or the planned next generation of collider experiments 
due to insufficient energy resolution. A numerically important but computationally challenging 
contribution to the off-shell regime is the interference contribution shown in 
Figure~\ref{fig:ggZZinterference} where the top-quark contribution to the $gg\to ZZ$ amplitude 
dominates the region of large invariant mass $M_{ZZ}$. The top-quark contribution has been computed 
at two-loop order within the LME \cite{Campbell:2016ivq,Caola:2016trd} and Pad\'e approximants 
based on the LME were studied in \cite{Campbell:2016ivq}. Only two form factors contribute to 
the Higgs-interference contribution and we show our approximate results based on the LME to $z^6$ 
and the threshold expansion up to at least $\bar{z}^{\,4}$ \cite{Grober:2019kuf} in Figure~\ref{fig:ggZZ}. 
Again we obtain an almost perfect reconstruction at LO. At NLO we get a very good prediction 
for the numerically dominant axial-vector form factor $\Ket{\widetilde{F}_{AA}^{(2)}}$ in the 
entire phase space. The errors are larger and the convergence is slower for the vector form factor 
$\Ket{\widetilde{F}_{VV}^{(2)}}$, because the contribution proportional to $\ln(-\hat{s}/m_t^2+i0)$ is 
larger relative to the non-logarithmic term and behaves worse as already mentioned for the case of 
Higgs-pair production. However, the discussion in \cite{Grober:2019kuf} demonstrates that the interference 
amplitude can be described reliably due to the smallness of the vector form factor. 

\section{Conclusions}

We have described our method for the reconstruction of top-quark mass effects in gluon fusion 
processes with a conformal mapping and Pad\'e approximants based on kinematic expansions. The 
reliability of the approach has been demonstrated with examples at two and three-loop order. 
While more numerical calculations have become available at two-loop order \cite{ProcRadcor,
Borowka:2016ehy,Borowka:2016ypz,Jones:2018hbb,Baglio:2018lrj,Chen:2019fla}, our method is the 
only approach beyond the LME that is feasible at three loops with the current computational 
technology and can significantly improve the quality of predictions at this order~\cite{ggHH3Loops}.

\end{document}